\documentclass{aa}
\usepackage{graphicx}
\def\deg{\ifmmode^{\circ}\;\else$^{\circ}\;$\fi} % overwrites \deg in LaTeX
\def\lsim{\,\lower2truept\hbox{${<\atop\hbox{\raise4truept\hbox{$\sim$}}}$}\,}
\def\gsim{\,\lower2truept\hbox{${> \atop\hbox{\raise4truept\hbox{$\sim$}}}$}\,}
\begin{document}

   \title{Predictions for high-frequency radio surveys of extragalactic sources}
%   \title{Periodic behaviors in the radio light curves of BL objects}
   \subtitle{}

   \author{Gianfranco De Zotti \inst{1,2}
        \and
        Roberto Ricci \inst{2}
           \and Dino Mesa \inst{1}
        \and Laura Silva \inst{3}
        \and Pasquale Mazzotta \inst{4}
        \and Luigi Toffolatti \inst{5} \and Joaqu{\'\i}n Gonz\'alez-Nuevo\inst{5} }

   \offprints{G. De Zotti: dezotti@pd.astro.it}

\institute{ INAF--Osservatorio Astronomico di Padova, Vicolo
dell'Osservatorio 5, I-35122 Padova,
   Italy\\ \email{dezotti@pd.astro.it}
   \and International School for Advanced Studies,
SISSA/ISAS, Via Beirut 2-4, I-34014 Trieste, Italy\\
         \email{ricci@sissa.it}
\and INAF--Osservatorio Astronomico di Trieste, Via G.B. Tiepolo
11, I-34131 Trieste, Italy \\ \email{silva@ts.astro.it}
 \and Departimento di Fisica, Universit\`a di Roma ``Tor Vergata", Via della Ricerca
Scientifica 1, I--00133 Roma Italy
\\\email{pasquale.mazzotta@roma2.infn.it} \and Departamento de F{\'\i}sica, Universidad de Oviedo, c.
Calvo Sotelo s/n, 33007 Oviedo, Spain }

   \date{Received ..... ; accepted ........}

   \abstract{We present detailed predictions of the contributions
   of the various source populations to the counts at frequencies
   of tens of GHz. New evolutionary models are worked out for
   flat-spectrum radio quasars, BL Lac objects, and
   steep-spectrum sources. Source populations characterized by spectra peaking at high
   radio frequencies, such as extreme GPS sources, ADAF/ADIOS sources and
   early phases of $\gamma$-ray burst afterglows are also dealt with.
   The counts of different populations of
   star-forming galaxies (normal spirals, starbursts, high-$z$
   galaxies detected by SCUBA and MAMBO surveys, interpreted as
   proto-spheroidal galaxies) are estimated taking into account
   both synchrotron and free-free emission, and dust re-radiation.
   Our analysis is completed by updated counts of Sunyaev-Zeldovich
   effects in clusters of galaxies and by a preliminary estimate
   of galactic-scale Sunyaev-Zeldovich signals associated to
   proto-galactic plasma.

    }
\titlerunning{High-frequency radio surveys}

  \maketitle

%
%________________________________________________________________

\section{Introduction}

The new broad-band correlators for compact interferometric arrays
and multi-beam receivers are  beginning to enable large area
surveys at high radio frequencies ($\ge 15\,$GHz), hitherto
impossible because the beam solid angle scales as $\nu^{-2}$
(assuming diffraction limited performance) so that surveying large
areas of the sky becomes quickly impractical at high frequencies
with conventional techniques. The high-frequency surveys will have
a major impact on astrophysics:

\begin{description}

\item[--]
They will open a window on new classes of both Galactic and
extragalactic sources, for example on those with strong
synchrotron or free-free self-absorption corresponding to both
very early phases of nuclear radio-activity (extreme GHz Peaked
Spectrum - GPS - sources or high-frequency peakers) and late
phases of the evolution of Active Galactic Nuclei (AGNs),
characterized by low accretion/radiative efficiency (ADAF/ADIOS
sources), as well as to early phases of the evolution of radio
afterglows of gamma-ray bursts.

\item[--] They will provide adequate samples to get a really unbiased
view of rare, but very interesting, classes of sources with flat
spectrum up to high frequencies, that, at low frequencies, are
swamped by more numerous populations which fade away as the
frequency increases. One example are blazars, for which much more
extended samples are required to properly cover their parameter
space. Galactic sources with substantial high-frequency emission
include star-forming regions, transient black hole binaries and
flaring brown dwarves.

\item[--] They will allow us to get important information even on the
physics of sources known from lower frequency surveys: the high
frequency spectral steepenings due to electron aging are
informative on the distribution of source ages and on mechanisms
for energy losses; the break frequencies marking the transition
from optically thick to optically thin regimes depend on the
magnetic field intensity, etc..

\item[--] They will produce surveys of the Sunyaev-Zeldovich effect
in distant clusters of galaxies and perhaps on galactic scales,
extremely important both to understand the formation of large
scale structure and the heating of the intergalactic medium.

\item[--] They will play a vital role in the interpretation of
temperature and polarization maps of the Cosmic Microwave
Background (CMB), by allowing us to characterize and remove the
contamination by astrophysical foregrounds

\item[--] They will be essential to calibrate the forthcoming large
millimeter array ALMA.

\item[--] Follow-up high frequency observations will allow us to
study the intrinsic polarization of sources, unaffected by Faraday
depolarization, as well as their variability properties, essential
ingredients to understand their physics.

\end{description}

\noindent NASA's WMAP satellite, designed to make accurate
measurements of the space distribution of the Cosmic Microwave
Background (CMB), is carrying out the first all sky millimeter
surveys in five bands (centered at 22.8, 33.0, 40.7, 60.8, and
93.5 GHz; Bennett et al. 2003a). The analysis of the first year
observations have yielded a 98\% reliable catalog of 208 point
sources brighter than $\sim 1\,$Jy (Bennett et al. 2003b).

Unbiased radio surveys at 15 GHz, covering $520\,\hbox{deg}^2$,
have been carried out with the Ryle telescope (Taylor et al. 2001;
Waldram et al. 2003). Ricci et al. (2004), using a novel wide-band
analogue correlator on one baseline of the Australia Telescope
Compact Array (ATCA) have surveyed $1216\,\hbox{deg}^2$ of the
southern sky at 18 GHz. Preliminary source counts at 30 GHz have
been produced by the DASI (Kovac et al. 2002), VSA (Grainge et al.
2003), and CBI (Mason et al. 2003) CMB experiments.

The ATCA survey will cover, in the near future,
$10^4\,\hbox{deg}^2$ at 20 GHz to a flux limit $S_{20{\rm GHz}} >
40\,$mJy. The Jodrell Bank Observatory -- University of Torun One
Centimetre Radio Array (OCRA) will begin soon a multi-beam survey
of the Northern sky at 30 GHz. ESA's {\sc Planck} satellite will
carry out all-sky surveys at 9 frequencies from 30 to 860 GHz with
sensitivity several times higher than WMAP.

As a new observational window opens up, it is useful to utilize
models exploiting the available information to foretell which
source populations the planned surveys will be able to detect as a
function of their depth, which redshift range will these survey
explore, which information on formation and evolution of each
source population will they be able to provide. Answering these
questions is a key step towards an optimal planning of the
surveys.

Earlier time-honored comprehensive analyses of radio-source
evolution (Dunlop \& Peacock 1990; Toffolatti et al. 1998; Jackson
\& Wall 1999) provided remarkably successful fits to data from
surveys at $\lsim 8\,$GHz, covering several decades in
flux-density. The model by Toffolatti et al. (1998), in
particular, was extensively exploited to estimate the radio source
contamination of CMB maps at cm/mm wavelengths (e.g.:
Mart{\'\i}nez-Gonz\'alez et al. 2003; Vielva et al. 2003; Herranz
et al. 2002; Refregier et al. 2000) and turns out to
satisfactorily account also for the high frequency counts down to
$S \simeq 20\,$mJy, with a maximum offset of a factor $\simeq 0.7$
but only at the brightest fluxes (cf. Fig. 13 of Bennett et al.
2003). On the other hand, the data that have been accumulating in
recent years require a more detailed treatment of each of the
various relevant sub-populations.

In this paper we present a new analysis encompassing canonical
radio source populations as well as a variety of special classes
of sources. In Section 2 we describe the model for canonical radio
sources, allowing for different evolutionary behaviours for two
classes of flat-spectrum sources (flat-spectrum radio quasars and
BL Lac objects) and for steep-spectrum sources, while in Sect. 3
we present the data sets used to determine the parameters. In
Sect. 4 we work out predictions of counts for star-forming
galaxies, extreme GPS sources, ADAF/ADIOS sources,
Sunyaev-Zeldovich effects on scales from clusters of galaxies to
large galaxies, radio afterglows of $\gamma$-ray bursts. Our main
conclusions are summarized and discussed in Sect.~5.

We have adopted a flat $\Lambda$CDM cosmology with
$\Omega_{\Lambda}=0.7$ and
$H_0=65\,\hbox{km}\,\hbox{s}^{-1}\,\hbox{Mpc}^{-1}$.

\begin{figure}
\begin{center}
\includegraphics[height=7cm, width=7cm]{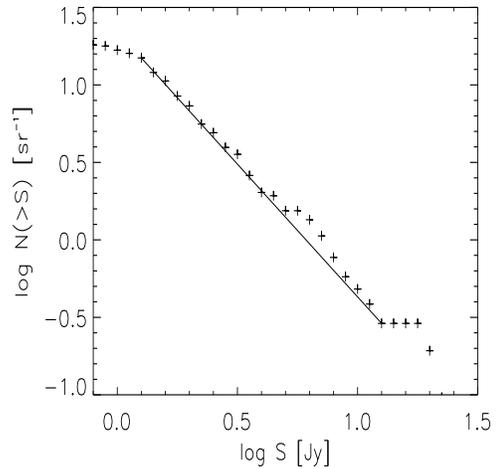}
\caption{Integral counts of extragalactic sources in the WMAP
$K$-band (centered at 22.8 GHz). }\label{WMAP_counts}
\end{center}
\end{figure}

\begin{figure}
\begin{center}
\includegraphics[height=7cm, width=7cm]{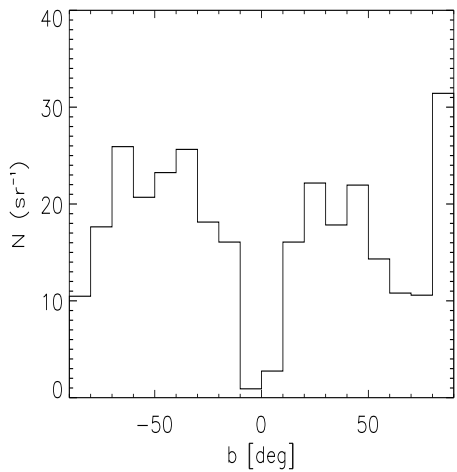}
\caption{Surface density of WMAP extragalactic sources as a
function of Galactic latitude. }\label{WMAP_lat}
\end{center}
\end{figure}

\begin{table*}
\begin{center}
\caption{Best fit values of the parameters of the evolutionary
models for canonical radio sources.}\label{tab:parameters}
\begin{tabular}{lcrrccrr}
  \hline
Source type &\multicolumn{4}{c}{luminosity function}&\ &
\multicolumn{2}{c}{evolution}\\ \cline{2-5} \cline{7-8}
 & \multicolumn{1}{c}{$\log n_0(\hbox{Mpc}^{-3})$}
 & \multicolumn{1}{c}{$a$}& \multicolumn{1}{c}{$b$}&
 \multicolumn{1}{c}{$\log
 L_\ast(\hbox{erg}\,\hbox{s}^{-1}\,\hbox{Hz}^{-1}@5\hbox{GHz},z=0)$} & \
 & \multicolumn{1}{c}{$k_{\rm ev}$} & \multicolumn{1}{c}{$z_{\rm top}$} \\
\hline
FSRQ    & -8.988 & 0.651 &  2.888 & 34.043 & & 0.224 & 2.252 \\
BL Lac  & -7.921 & 0.977 &  1.266 & 32.789 & & 0.745 &       \\
Steep   & -8.030 & 1.071 &  3.054 & 33.329 & & 4.902 &       \\
 \hline
\end{tabular}
\end{center}
\end{table*}

\section{The evolutionary model for canonical radio source populations}

We have considered different epoch-dependent luminosity functions
(LFs) for two flat-spectrum [flat-spectrum radio quasars (FSRQs)
and BL Lacertae type objects (BL Lacs)], and for steep-spectrum
radio source populations. The study by Sadler et al. (2002) of the
local radio luminosity function at 1.4 GHz yielded separate
estimates for sources powered by nuclear activity (AGN) and for
star-forming galaxies. By ``steep-spectrum sources'' we mean here
only the former population; the latter is considered in
Sect.~\ref{starforming}.

We have adopted LFs (in units of $\hbox{Mpc}^{-3}\,(d\log
L)^{-1}$) of the form
\begin{equation}
\Phi(L,z)=\frac{n_0}{(L/L_\ast)^{a}+(L/L_\ast)^{b}} \ .
\label{lum_func1}
\end{equation}
%
%The exception are FRI sources for which we have adopted the
%analytic expression given by Dunlop \& Peacock (1990) for
%low-power steep-spectrum sources (their Eq.~(9)).
%
%Luminosity evolution was assumed for all sub-populations, except
%for FRIs which, following Dunlop \& Peacock (1990), were assumed
%not to evolve.
Only in the case of FSRQs enough information is available to look
(although in a simplified manner) for evidences of a decline of
the space density at high redshifts. For these sources, we have
assumed:
\begin{equation}
L_{\ast,{\rm FSRQ}}(z)=L_\ast(0) 10^{k_{\rm ev}z(2z_{\rm top}-z)}
\ . \label{lum_funcBLL }
\end{equation}
For the other evolving populations we have used the simpler
expression:
\begin{equation}
L_\ast(z)=L_\ast(0) \exp[k_{\rm ev}\tau(z)] \ , \label{Last}
\end{equation}
where $\tau(z)$ is the look-back time in units of the Hubble time,
$H_0^{-1}$. No evolution was assumed for the lowest luminosity
steep-spectrum sources ($\log
L_{5\rm{GHz}}(\hbox{erg}\,\hbox{s}^{-1}\,\hbox{Hz}^{-1}) < 30.5$).

We have allowed for the high-frequency spectral steepenings of
steep-spectrum radio sources exploiting the quadratic fits by
FR-type and radio power derived by Jackson \& Wall (2001; their
Table 2). Simple power-law spectra with index $\alpha_{\rm
flat}=-0.1$ ($S_\nu \propto \nu^\alpha$), consistent with the
results by Ricci et al. (2004), have been adopted for
flat-spectrum sources.

The best fit values of the parameters  were derived using the
routine ``amoeba" (Press et al. 1992) exploiting the downhill
simplex method in multidimensions. The fitted data sets are
described below and the parameter values are listed in
Table~\ref{tab:parameters}.

\begin{figure}
\begin{center}
\includegraphics[height=7cm, width=7cm]{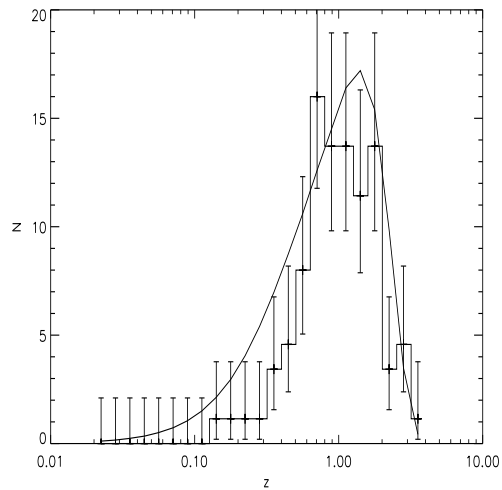}
\caption{Redshift distribution of WMAP FSRQs (histogram) compared
with the best-fit model (solid line). }\label{WMAP_zdistr_FSRQ}
\end{center}
\end{figure}

%\begin{figure}
%\begin{center}
%\includegraphics[height=7cm, width=7cm]{WMAP_zdistr_BLLac.ps}
%\caption{Redshift distribution of WMAP BL Lacs (histogram)
%compared with the best-fit model (solid line).
%}\label{{WMAP_zdistr_BLLac}}
%\end{center}
%\end{figure}

\begin{figure}
\begin{center}
\includegraphics[height=7cm, width=7cm]{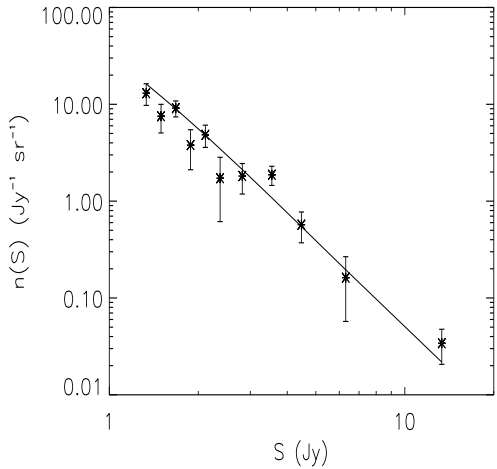}
\caption{Differential counts of WMAP FSRQs compared with the
best-fit model (solid line). }\label{WMAP_counts_FSRQ}
\end{center}
\end{figure}

\begin{figure}
\begin{center}
\includegraphics[height=7cm, width=7cm]{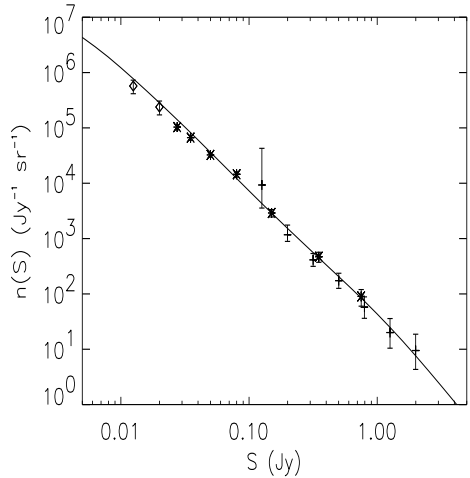}
\caption{Comparison of the model (solid line) with the
differential 15~GHz counts from the main survey (filled circles)
and the deeper survey (filled squares) by Waldram et al.~(2003)
and with total differential counts at 18 GHz by Ricci et al.
(2004; crosses). }\label{PS18_Waldram_counts}
\end{center}
\end{figure}

\section{Data sets}

\subsection{The first year WMAP catalog of extragalactic sources}

As illustrated by Fig.~\ref{WMAP_counts}, the first year WMAP
catalog of extragalactic sources (Bennett et al. 2003) appears to
be complete to $S_K \simeq 1.25\,$Jy in the $K$-band, centered at
22.8 GHz (see also Arg\"ueso et al. 2003). Their distribution as a
function of Galactic latitude (Fig.~\ref{WMAP_lat}) shows a clear
deficit at $|b| < 10^\circ$. Removing the 4 sources in this
region, we are left with 155 sources above the completeness limit,
over an area of 10.4 sr. For all these sources we have determined
the 4.85--22.8 GHz spectral index, using the 4.85 flux densities
from the GB6 (Gregory et al. 1996) or PMN (available at
http://www.parkes.atnf.csiro.au/) catalogs. We have searched the
NED and VizieR on-line databases for optical identifications and
redshifts. The search yielded 91 flat-spectrum radio quasars
(FSRQs), 28 BL Lacs, 17 unclassified flat-spectrum sources and 19
steep-spectrum sources (we define as flat-spectrum sources those
with spectral index $\alpha > -0.5$, $S_\nu \propto \nu^\alpha$).
Redshift measurements were found for 86 FSRQs, 27 BL Lacs, 9 of
the unclassified sources flat-spectrum sources and for 18
steep-spectrum sources. The redshift distributions and the counts
of FSRQs (Figs.~\ref{WMAP_zdistr_FSRQ} and
~\ref{WMAP_counts_FSRQ}) and of BL Lacs have been corrected by a
factor of $(1+17/119)$ to include the 17 unclassified
flat-spectrum sources, partitioned among the two populations in
proportion to the number of classified sources in each population.
In computing the model redshift distributions, the surveyed area
has been decreased by a factor of $0.94$ in the case of FSRQs and
of $0.96$ in the case of BL Lacs, to allow for the incompleteness
of redshift measurements. Poisson errors (Gehrels 1986) have been
adopted both for the counts and the redshift distributions.

\begin{figure}
\begin{center}
\includegraphics[height=7cm, width=7cm]{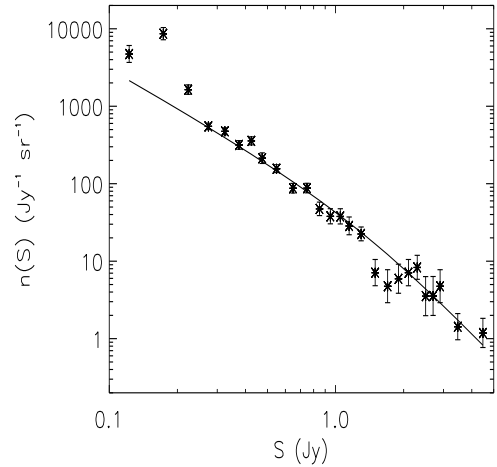}
\caption{Counts of FSRQs in the Parkes quarter-Jy sample compared
with the best-fit model (solid line). }\label{qJy_counts_FSRQ}
\end{center}
\end{figure}

\begin{figure}
\begin{center}
\includegraphics[height=7cm, width=7cm]{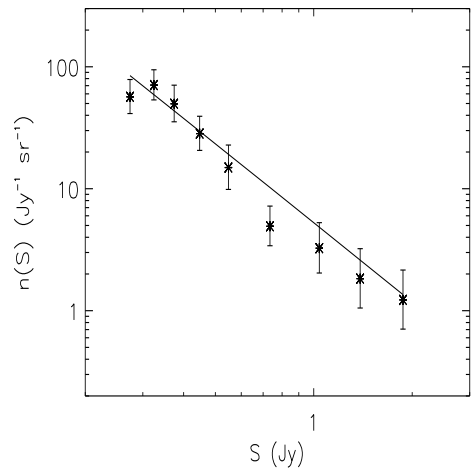}
\caption{Counts of BL Lac objects in the Parkes quarter-Jy sample
compared with the best-fit model (solid line).
}\label{qJy_counts_BLL}
\end{center}
\end{figure}

\begin{figure}
\begin{center}
\includegraphics[height=7cm, width=7cm]{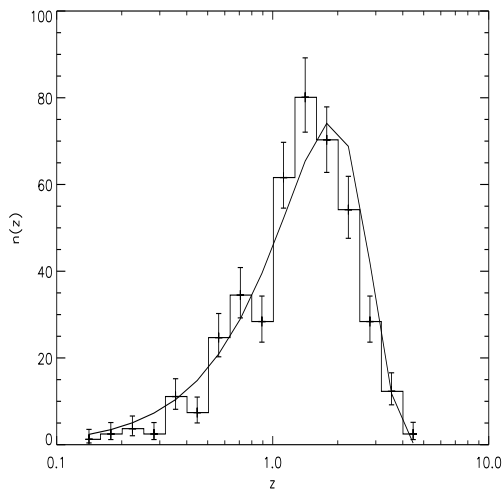}
\caption{Redshift distribution of FSRQs in the Parkes quarter-Jy
sample (histogram) compared with the best-fit model (solid line).
}\label{qJy_z_FSRQ}
\end{center}
\end{figure}

\subsection{The Parkes quarter-Jansky flat-spectrum sample}

The sample (Jackson et al. 2002) comprises 878 sources with
spectral index $\alpha^{5{\rm GHz}}_{2.7{\rm GHz}} \le -0.4$,
selected from several surveys to different depths at 2.7 GHz. The
catalog compiled by Jackson et al. (2002) includes optical
identifications and redshifts for most of the sources; 827 are
compact extragalactic radio sources, 38 have either extended radio
structure or are Galactic objects, 13 are unidentified (4 of which
because of obscuration by Galactic stars). Of the 827 compact
extragalactic radio sources, 624 are FSRQs, 54 are BL Lacs and 149
flat-spectrum radio sources. As in the case of WMAP data, we have
subdivided the latter plus the unidentified sources among the
FSRQs and the BL Lacs, in proportion to the fraction of classified
sources belonging to each population, i.e. we increase the counts
of both populations by a factor $1+162/(624+54)$.

While the full sample has been used to derive the differential
source counts (shown in Figs.~\ref{qJy_counts_FSRQ} and
\ref{qJy_counts_BLL}), for the analysis of the redshift
distribution we have defined a complete sub-sample aiming at
maximizing the fraction of sources with measured redshift. To this
end, we have first focussed on areas surveyed to at least 0.25 Jy,
keeping only sources brighter than this limit, for a total of 618
sources. Of these, 437 are FSRQs, 360 of which (82\%) have
measured redshift, 47 are BL Lacs, only 10 of which (21\%) have
measured redshift, and 111 are flat-spectrum galaxies (32, i.e.
29\%, with redshift). Clearly, only in the case of FSRQs the
fraction of sources with measured redshifts is large enough for a
meaningful redshift distribution to be derived. Further
restricting the sample to $\delta >-50^{\deg}$, we are left with
370 FSRQs, 345 of which (93\%) with measured redshift, 47 BL Lacs,
95 sources classified as galaxies and 2 unidentified sources. The
redshift distribution of FSRQs for this sub-sample is thus well
defined (Fig.~\ref{qJy_z_FSRQ}). As in the case of the counts, we
have corrected the redshift distribution by a factor of
$1+(95+2)/(370+47)$ to allow for a fraction of flat-spectrum
galaxies and of unidentified sources proportional to the ratio
FSRQ/(FSRQ$+$BL Lacs).

\begin{figure}
\begin{center}
\includegraphics[height=7cm, width=7cm]{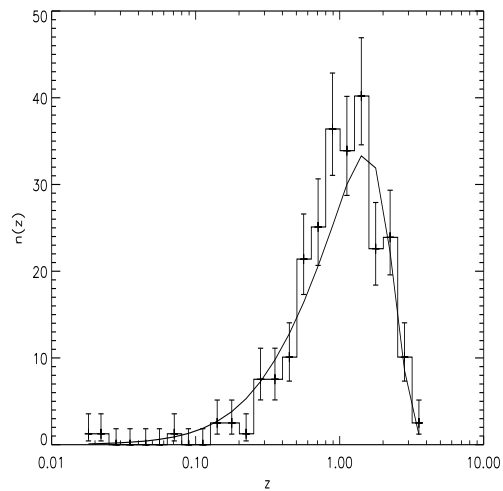}
\caption{Redshift distribution of FSRQs in the K\"uhr et al.
(1981) sample (histogram) compared with the best-fit model (solid
line). }\label{KUHR_z_FSRQ}
\end{center}
\end{figure}

%\begin{figure}
%\begin{center}
%\includegraphics[height=7cm, width=7cm]{KUHR_z_BLL.ps}
%\caption{Redshift distribution of BL Lacs in the K\"uhr et al.
%(1981) sample (histogram) compared with the best-fit model (solid
%line). }\label{KUHR_z_BLL}
%\end{center}
%\end{figure}

\subsection{The K\"uhr 1 Jy sample}

The sample (K\"uhr et al. 1981) comprises 518 sources to a 5 GHz
flux density limit of 1 Jy, over an area of 9.811 sr. Based on the
catalogued spectral indices, 299 sources are flat-spectrum
($\alpha > -0.5$, $S_\nu \propto \nu^\alpha$) and 219 are
steep-spectrum. The former population includes 212 FSRQs, 200 of
which (94\%) have measured redshift, 26 BL Lacs (20 of which,
77\%, with measured redshift, and 61 either classified as galaxies
or missing a morphological classification. As before, we have
distributed the latter sources among FSRQs and BL Lacs, thus
increasing the counts of both types by a factor of
$1+61/(212+26)$. The fit of the redshift distribution of FSRQs is
shown in Fig.~\ref{KUHR_z_FSRQ}. %The steep-spectrum sources, 178
%of which (81\%) have measured redshift, have all been taken as
%FRII.

\begin{figure}
\begin{center}
\includegraphics[height=7cm, width=7cm]{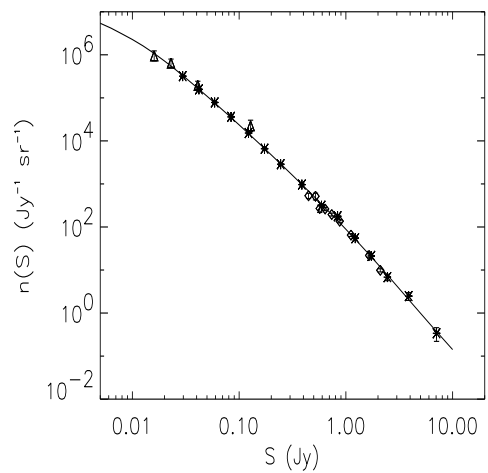}
\caption{Differential counts at 5 GHz compared with the best-fit
model (solid line). Data from Gregory \& Condon (1991, asterisks)
and Pauliny-Toth et al. (1978a, diamonds; 1978b, triangles).
}\label{5GHz_counts}
\end{center}
\end{figure}

\begin{figure}
\begin{center}
\includegraphics[height=7cm, width=7cm]{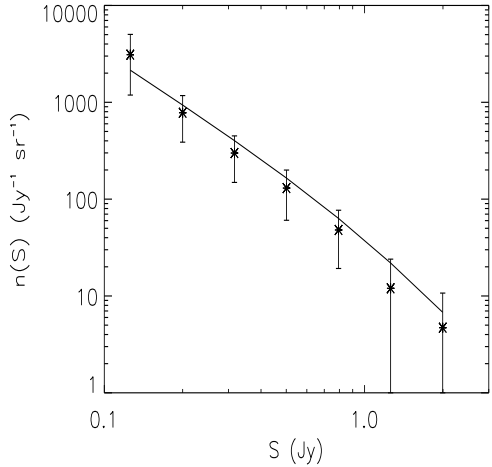}
\caption{Differential counts at 18 GHz of flat-spectrum sources by
Ricci et al. (2004) compared with the best-fit model (solid line).
} \label{PS18_flat_counts}
\end{center}
\end{figure}

\begin{figure}
\begin{center}
\includegraphics[height=7cm, width=7cm]{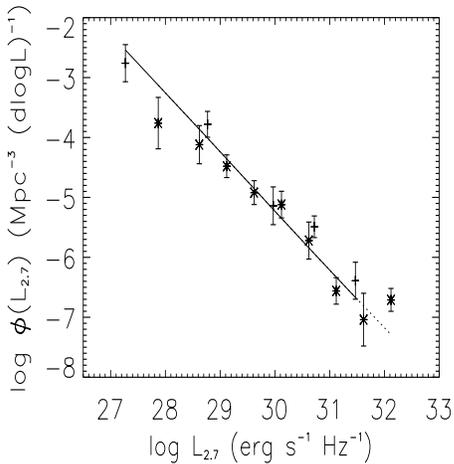}
\caption{Local luminosity functions at 2.7 GHz by Peacock (1985)
(crosses) and Toffolatti et al. (1987) (asterisks) compared with
the best-fit model (solid line).} \label{flatFL}
\end{center}
\end{figure}

\subsection{Other data sets}

Furthermore, we have fitted the 5~GHz counts
(Fig.~\ref{5GHz_counts}) by Gregory \& Condon (1991) and
Pauliny-Toth et al. (1978a,b), the 15 GHz 9C counts plus the 18
GHz ATCA counts (Fig.~\ref{PS18_Waldram_counts}) by Waldram et al.
(2003) and Ricci et al. (2004), respectively, and the ATCA 18 GHz
counts (Fig.~\ref{PS18_flat_counts}) of flat-spectrum sources
(Ricci et al. 2004).

In our scheme, blazars (FSRQs plus BL Lacs) account for the entire
population of flat-spectrum radio sources. We thus require that
the sum of their local luminosity functions matches the estimates
of the local luminosity function of flat-spectrum sources by
Peacock (1985) and Toffolatti et al. (1987). The fit is shown in
Fig.~\ref{flatFL}.

\section{Special source populations}

\subsection{Star-forming galaxies}\label{starforming}

The radio emission of galaxies correlates with their
star-formation rate, as demonstrated by the well established tight
correlation with far-IR emission (Helou et al. 1985; Gavazzi et
al. 1986; Condon 1992; Garrett 2002; Morganti et al. 2004). The
cross-over between synchrotron plus free-free emission, prevailing
at cm wavelengths, and thermal dust emission generally occurs at
$\lambda \simeq 2$--3 mm, so that at frequencies of tens of GHz
there are contributions from both components, the former being
associated to normal late-type and starburst galaxies at low to
moderate redshifts, the latter to the high-redshift population
detected by sub-mm surveys (``SCUBA galaxies'').

Following Granato et al. (2001, 2004) we interpret ``SCUBA
galaxies'' as proto-spheroidal galaxies in the process of forming
most of their stars. Their epoch-dependent luminosity function is
computed using the physical model by Granato et al. (2004), which
describes the star-formation and chemical enrichment histories
and, interfaced with the code GRASIL (Silva et al. 1998), yields
the spectrophotometric evolution from radio to X-ray wavelengths.
The values of GRASIL parameters were determined by Silva et al.
(2004) comparing the model predictions with a broad variety of
observational data.

Normal late-type and starburst galaxies are dealt with in a more
phenomenological manner, following Silva et al. (2004). Briefly,
we adopted the $60\,\mu$m local luminosity functions  of ``warm''
and ``cold'' galaxies (based on IRAS colors), determined by
Saunders et al. (1990), for starburst and late-type galaxies,
respectively. The local luminosity functions were cut off at
$L_{60\mu{\rm m}} = 2\times 10^{32}$ erg/s/Hz. Extrapolations to
other wavelengths were made using GRASIL spectral energy
distributions fitting the data on NGC$\,6090$ (for starburst
galaxies) and NGC$\,6946$ (for late-type galaxies).

A density and luminosity evolution model ($LF[L(z),z] =
LF[L(z)/(1+z)^{2.5}, z=0]\times (1+z)^{3.5}$) was adopted for
starburst galaxies, while a mild pure luminosity evolution
($L(z)=L(0)\,(1+z)^{1.5}$) was assumed for normal late-type
galaxies. These evolutionary laws were applied up to $z=1$. The
luminosity functions were then kept to the (comoving) values they
have at $z=1$ up to a redshift cut-off $z_{\rm cutoff} =1.5$.

\subsection{Extreme GHz peaked spectrum (GPS) sources}

A class of sources that is expected to come out in high frequency
surveys is that of extreme GHz Peaked Spectrum (GPS) or  very high
frequency peakers. GPS sources are powerful ($\log P_{\rm 1.4\,
GHz} \gsim 25\,\hbox{W}\,\hbox{Hz}^{-1}$), compact ($\lsim
1\,$kpc) radio sources with a convex spectrum peaking at GHz
frequencies (see O'Dea 1998 for a comprehensive review). They are
identified with both galaxies and quasars.

It is now widely agreed that GPS sources correspond to the early
stages of the evolution of powerful radio sources, when the radio
emitting region grows and expands within the interstellar medium
of the host galaxy, before plunging in the intergalactic medium
and becoming an extended radio source (Fanti et al. 1995; Readhead
et al.  1996; Begelman 1996; Snellen et al. 2000). Conclusive
evidence that these sources are young came from measurements of
propagation velocities. Velocities of up to $\simeq 0.4c$ were
measured, implying dynamical ages $\sim 10^3$ years (Polatidis et
al. 1999; Taylor et al. 2000; Tschager et al. 2000). The
identification and investigation of these sources is therefore a
key element in the study of the early evolution of radio-loud
AGNs.

There is a clear anti-correlation between the peak (turnover)
frequency and the projected linear size of GPS sources (and of
compact steep-spectrum sources), suggesting that the process
(probably synchrotron self-absorption, although free-free
absorption is also a possibility) responsible for the turnover
depends simply on the source size. Although this anti-correlation
does not necessarily define the evolutionary track, a decrease of
the peak frequency as the emitting blob expands is indicated. Thus
high-frequency surveys may be able to detect these sources very
close to the moment when they turn on.

The self-similar evolution models by Fanti et al. (1995) and
Begelman (1996) imply that the radio power drops as the source
expands, so that GPS's evolve into lower luminosity radio sources,
while their luminosities are expected to be very high during the
earliest evolutionary phases, when they peak at high frequencies.
De Zotti et al. (2000) showed that, with a suitable choice of the
parameters, this kind of models may account for the observed
counts, redshift and peak frequency distributions of the samples
then available. The data indicated strongly different evolutionary
properties between GPS galaxies and quasars, at variance with
unification scenarios.

The models by De Zotti et al. (2000) imply that extreme GPS
quasars, peaking at $\nu > 20\,$GHz, comprise a quite substantial
fraction of bright radio sources in the WMAP survey at $\nu \simeq
20\,$GHz, while GPS galaxies with similar $\nu_{\rm peak}$ should
be about 10 times less numerous. For a maximum rest-frame peak
frequency $\nu_{p,i} =200\,$GHz, the model predicts about 10 GPS
quasars with $S_{30{\rm GHz}} > 2\,$Jy peaking at $\geq 30\,$GHz
over the 10.4 sr at $|b| >10^\circ$.

Although the number of {\it candidate} GPS quasars in the WMAP
survey is consistent with that expected from the models, when data
at additional frequencies (Trushkin 2003) are taken into account
such candidates look more blazars caught during a flare optically
thick up to high frequencies. Furthermore, Tinti et al. (2004)
have shown that most, perhaps two thirds, of the quasars in the
sample of High Frequency Peaker candidates selected by Dallacasa
et al. (2000) are likely blazars, while 9 out of 10 candidates
classified as galaxies are consistent with being truly young
sources.

Snellen et al. (2000) support a scenario whereby the luminosity of
GPS sources {\it increases} with time (while the peak frequency
decreases) until a linear size $\simeq 1\,$kpc is reached, and
decreases subsequently. Although predictions for GPS source counts
have not been worked out in this framework, these objects must be
very rare at bright high frequency fluxes. In
Figs.~\ref{tot_counts20} and \ref{tot_counts30} we have
provisionally plotted (dashed line in the upper right-hand panel)
the predictions of the models by De Zotti et al. (2000) with a
maximum intrinsic peak frequency $\nu_{p,i} = 200\,$GHz,
decreasing by a factor of 3 the contribution of GPS quasars.

Large area high frequency surveys, coupled with follow-up
simultaneous  multifrequency observations of GPS candidates, would
be essential to assess the evolutionary properties of this
population.

\subsection{Late stages of AGN evolution}

High radio frequency observations are also crucial to investigate
late stages of the AGN evolution, characterized by low
radiation/accretion efficiency. This matter was recently brought
into sharper focus by the discovery of ubiquitous, moderate
luminosity  hard X-ray emission from nearby ellipticals. VLA
studies at high radio frequencies (up to 43 GHz) have shown,
albeit for a limited sample of objects, that all the observed
compact cores of elliptical and S0 galaxies have spectra rising up
to $\simeq 20$--30 GHz (Di Matteo et al. 1999).

There is growing evidence that essentially all massive ellipticals
host super-massive black holes (see, e.g., Kormendy \& Gebhardt
2001). Yet, nuclear activity is not observed at the level expected
from Bondi's (1952) spherical accretion theory, in the presence of
extensive hot gaseous halos, and for the usually assumed radiative
efficiency $\simeq 10\%$ (Di Matteo et al. 1999). However, as
proposed by Rees et al. (1982), the final stages of accretion in
elliptical galaxies may occur via Advection-Dominated Accretion
Flows (ADAFs), characterized by a very low radiative efficiency
(Fabian \& Rees 1995). The ADAF scenario implies strongly
self-absorbed thermal cyclo-synchrotron emission due to a near
equipartition magnetic field in the inner parts of the accretion
flows, most easily detected at cm to mm wavelengths. However the
ADAF scenario is not the only possible explanation of the data,
and is not problem-free. {\it Chandra} X-ray observations of Sgr
A, at the Galactic Center, are suggestive of a considerably lower,
compared to Bondi's, accretion rate (Baganoff et al. 2003), so
that the very low ADAF radiative efficiency may not be required. A
stronger argument against a pure ADAF in the Galactic Center is
that the emission is strongly polarized at mm/sub-mm wavelengths
(Aitken et al. 2000; Agol 2000). Also Di Matteo et al. (1999,
2001) found that the high frequency nuclear radio emission of a
number of nearby early-type galaxies is substantially below the
predictions of standard ADAF models. The observations would be
more consistent with the adiabatic inflow-outflow solutions
(ADIOS), developed by Blandford \& Begelman (1999), whereby the
energy liberated by the accretion drives an outflow at the polar
region carrying a considerable fraction of the mass, energy and
angular momentum available in the accretion flow, thus suppressing
the radio emission from the inner regions. Both the intensity and
the peak of the radio emission depend on the mass loss rate.

We have estimated the counts yielded by these low radiative
efficiency flows (LREF) following Perna \& Di Matteo (2000), i.e.
we have taken the number density of such sources to be given, as a
function of redshift, by their Eq.~(5), with a cut-off at $z=1$,
and have used the emission spectrum corresponding to $p=0.2$ in
their Fig.~1, which corresponds to the upper end of the range of
observed luminosities. The counts we obtain, shown by the dotted
line in the upper right-hand panel of Figs.~\ref{tot_counts20} and
\ref{tot_counts30}, are about a factor of ten below the estimates
by Perna \& Di Matteo (2000). We do not understand the reason for
this strong discrepancy. We note however, that the conclusion of
these Authors that the 30 GHz LREF counts can be comparable or
even higher than those of sources known from low-frequency surveys
is ruled out by the results of the Ryle telescope, ATCA and WMAP
surveys.

\section{Sunyaev-Zeldovich effects}

The  Sunyaev \& Zel'dovich (SZ) effect (Sunyaev \& Zeldovich 1972)
arises due to the inverse Compton scatter of CMB photons against
the hot electrons. The CMB intensity change is given by:
\begin{equation}
\Delta I_\nu = 2{(k T_{\rm CMB})^3 \over (hc)^2 }y g(x)
\end{equation}
where $T_{\rm CMB}$ is the CMB temperature and $x=h\nu/kT_{\rm
CMB}$.

The spectral form of this ``thermal effect" is described by the
function
\begin{equation}
g(x) = x^4\hbox{e}^x [x\cdot \coth(x/2) -4]/(\hbox{e}^x -1)^2,
\end{equation}
which is negative (positive) at values of $x$ smaller (larger)
than $x_0=3.83$, corresponding to a critical frequency $\nu_0=217$
GHz.

The Comptonization parameter is
\begin{equation}
y = \int {kT_e  \over mc^2} n_e \sigma _T dl,
\end{equation}
where $n_e$ and $T_e $ are the electron density and temperature,
respectively, $\sigma _T$  is the Thomson cross section, and the
integral is over a line of sight  through the plasma.

With respect to the incident radiation field, the change of the
CMB intensity  across a cluster can be viewed as a net flux
emanating from the plasma cloud, given by the integral of
intensity change over the solid angle subtended by the cloud:
\begin{equation}
{\Delta F}_\nu  =\int  \Delta I_{\nu}d\Omega \propto Y\equiv \int
y d\Omega \label{eq:ly}
\end{equation}
In the case of hot gas trapped in the potential well due to an
object of total mass $M$, the parameter $Y$ in Eq.~(\ref{eq:ly}),
called integrated Y-flux, is proportional to the gas-mass-weighted
electron temperature $\langle T_e \rangle$ and to the gas mass
$M_g=f_gM$:
\begin{equation}
Y\propto  f_g \langle T_e \rangle M\ .
\end{equation}
At the frequencies considered here, the Y-flux is negative and can
therefore be distinguished from the positive signals due to the
other source populations.

\subsection{Sunyaev-Zeldovich effects in galaxy clusters}

The SZ effect from the hot gas responsible for the X-ray emission
of rich clusters of galaxies has been detected with high signal to
noise and even imaged in several tens of objects (Carlstrom et al.
2002).

Recent X-ray {\it Chandra} observations show that for massive
clusters the gas fraction is remarkably constant, $f_g=0.113\pm
0.005$, independently of the cluster redshift (Allen et al. 2002).
Simple scaling laws are used to derive, under the hydrostatic
equilibrium hypothesis, the relation between the mass and the
temperature of a cluster (see, e.g., Barbosa et al. 1996;
Colafrancesco et al. 1997; Bryan \& Norman 1998, and references
therein):
\begin{equation}
T(M,z)= T_{15} h^{2/3}\Biggr[{\Omega_0\Delta_c(\Omega_0,z)\over
180}\Biggl]^{1/3} \Biggr({M\over M_{15}}\Biggl)^{2/3}\!\!\!\!\!
(1+z), \label{m-t}
\end{equation}
where $\Delta_c$ is the non-linear  density contrast of a cluster
that collapsed at redshift $z$ and  $T_{15}$ is the temperature of
a cluster of mass $M_{15}=10^{15}h^{-1}M_\odot$ which collapses
today. We  calculate  $\Delta_c$ using the equations reported in
the appendix of  Colafrancesco et al. (1997): the $M-T$ relation
calculated in this way is perfectly consistent with Eq.~(9) of
Bryan \& Norman (1998). Both simulations and observations indicate
that temperatures of massive clusters follow quite well the
relationship with mass and redshift expressed by Eq.~(\ref{m-t})
(see, e.g., Ettori et al. 2004 and references therein).
Nevertheless, there is still some discussion on the exact value of
$T_{15}$, whose values range from 6.6~keV to 8.8~keV (see, e.g.,
Pierpaoli et al. 2001, and references therein). This represents,
at the moment, the largest source of uncertainty for the
normalization of the power spectrum of the matter density
fluctuation $P(k)\equiv |\delta_k|^2$ inferred from the observed
cluster X-ray temperature function (XTF). Such normalization is
usually given in term of the value of $\sigma_8$, i.e. the
variance of the density perturbation $\sigma^2(R,z)\propto
\int_o^{\infty}  k^3P(k,z) W^2(kR)dk/k$, where $W(kR)$ is the
windows function corresponding to the smoothing of the density
field (see, e.g., Peebles 1993), on the scale of $R=8h^{-1}{\rm
  Mpc}$. Lower values of $T_{15}$ result in higher
values of $\sigma_8$ and viceversa (see, e.g., Ikebe et al. 2002;
Pierpaoli et al. 2003). Here we choose $T_{15}=7.75\,$keV,
consistent with the simulations of Eke et al. (1998), and
$\sigma_8=0.85$ yielding an XTF consistent with the observed one
(see, e.g., Pierpaoli et al. 2003). The number counts for SZ
clusters are then given by:
\begin{equation}
 N(>\overline {\Delta F_\nu}) = \int {dV \over dz} dz
\int_{\overline {M} (\overline {\Delta F_\nu},z)} dM N(M,z),
%,\eqno(5)
\end{equation}
where $N(M,z)$ is the cluster mass  function and the lower bound
of the mass integral is determined by requiring that clusters of
mass $M>\overline M$ at redshift $z$ have  SZ fluxes    greater
than ${\overline {\Delta F_{\nu}}}$ (for details see, e.g.,
Colafrancesco et al. 1997). Here we adopt the mass function of
Sheth et al. (2001):
\begin{eqnarray}
N(M,z)dM &=&{ \sqrt{2 \, a \, A^2\over \pi}}\rho_0 { 1\over
M^2}\left[
  1+\left( \frac{ \sigma(M,z)}{\sqrt{a}\delta_c}
\right)^{2p} \right] \cdot \nonumber \\
&\cdot& { d \ln \sigma \over
  d\ln M } { \delta_c \over \sigma(M,z)} \exp\left[ - {1 \over 2}{
   a \delta_c^2\over \sigma^2(M,z)}   \right],
 \label{eq:ps}
\end{eqnarray}
where $\rho_0$ is the present-day mass density of the Universe,
$\delta_c=1.68$, $a=0.707$, $A=0.3222$, $p=0.3$.

We caution that the 20 and 30 GHz counts of cluster SZ effects
shown in the lower right-hand panel of Figs.~\ref{tot_counts20}
and \ref{tot_counts30} could be substantially affected by
radio-source contamination since radio sources are strongly
correlated with clusters of galaxies (Holder 2002; Massardi \& De
Zotti 2004). Although the available data do not allow a
quantitative estimate of the effect on counts, we note that a
single powerful radio source can fill in the SZ decrement of a
small group of galaxies.

\subsection{Galaxy-scale Sunyaev-Zeldovich effects}

Evidences of statistically significant detections at 30 GHz of
arcminute scale fluctuations well in excess of predictions for
primordial anisotropies of the cosmic microwave background (CMB)
have recently been obtained by the CBI (Mason et al. 2003) and
BIMA (Dawson et al. 2002) experiments. The interpretation of these
results is still debated. Extragalactic sources are potentially
the dominant contributor to fluctuations on these scales and must
be carefully subtracted out. And indeed both groups devoted a
considerable effort for this purpose. However a quite substantial
residual contribution to the CBI signal is difficult to rule out
(De Zotti et al. 2004; Holder 2002), while the residual radio
source contamination of the BIMA results is likely to be very
small.

If indeed the detected signal cannot be attributed to
extragalactic radio sources, its most likely source is the thermal
Sunyaev-Zeldovich (SZ) effect (Gnedin \& Jaffe 2001). The
small-scale fluctuations due to the SZ within rich clusters of
galaxies has been extensively investigated (Komatsu \& Kitayama
1999; Bond et al. 2002). The estimated power spectrum was found to
be very sensitive to the normalization ($\sigma_8$) of the density
perturbation spectrum. A $\sigma_8\gsim 1$, somewhat higher than
indicated by other data sets, is apparently required to account
for the CBI data.

On the other hand, significant SZ signals may be associated to the
formation of spheroidal galaxies (De Zotti et al. 2004;
Rosa-Gonzalez et al. 2004). The proto-galactic gas is expected to
have a large thermal energy content, leading to a detectable SZ
signal, both when the proto-galaxy collapses with the gas
shock-heated to the virial temperature (Rees \& Ostriker 1977;
White \& Rees 1978), and in a later phase as the result of strong
feedback from a flaring active nucleus (Ikeuchi 1981; Natarajan et
al. 1998; Natarajan \& Sigurdssson 1999; Aghanim et al. 2000;
Platania et al. 2002; Lapi et al. 2003).

These SZ signals, showing up on sub-arcmin scales, are potentially
able to account for the BIMA results. Proto-galactic gas heated at
the virial temperature and with a cooling time comparable with the
expansion time may provide the dominant contribution. It must be
stressed, however, that estimates are plagued by large
uncertainties reflecting our poor understanding of the complex
physics governing the thermal history of the proto-galactic gas,
and, as stressed by the authors themselves, the counts of galactic
scale SZ effects computed by De Zotti et al. (2004) may be more
conservatively viewed as upper limits. Furthermore, the radio
emission from the galaxy can partially fill in the SZ dip,
particularly if the galaxy hosts a radio active nucleus (Platania
et al. 2002).

On the other hand, the SZ effect turns out to be an effective
probe of the thermal state of the gas and of its evolution, so
that its detection, that may be within reach of future high
frequency surveys (cf. the lower right-hand panel of
Figs.~\ref{tot_counts20} and \ref{tot_counts30}) would provide
unique information on early phases of galaxy evolution,
essentially unaccessible by other means.

\subsection{Radio afterglows of $\gamma$-ray bursts}

\begin{figure*}
\begin{center}
\includegraphics[height=11.5cm, width=14cm]{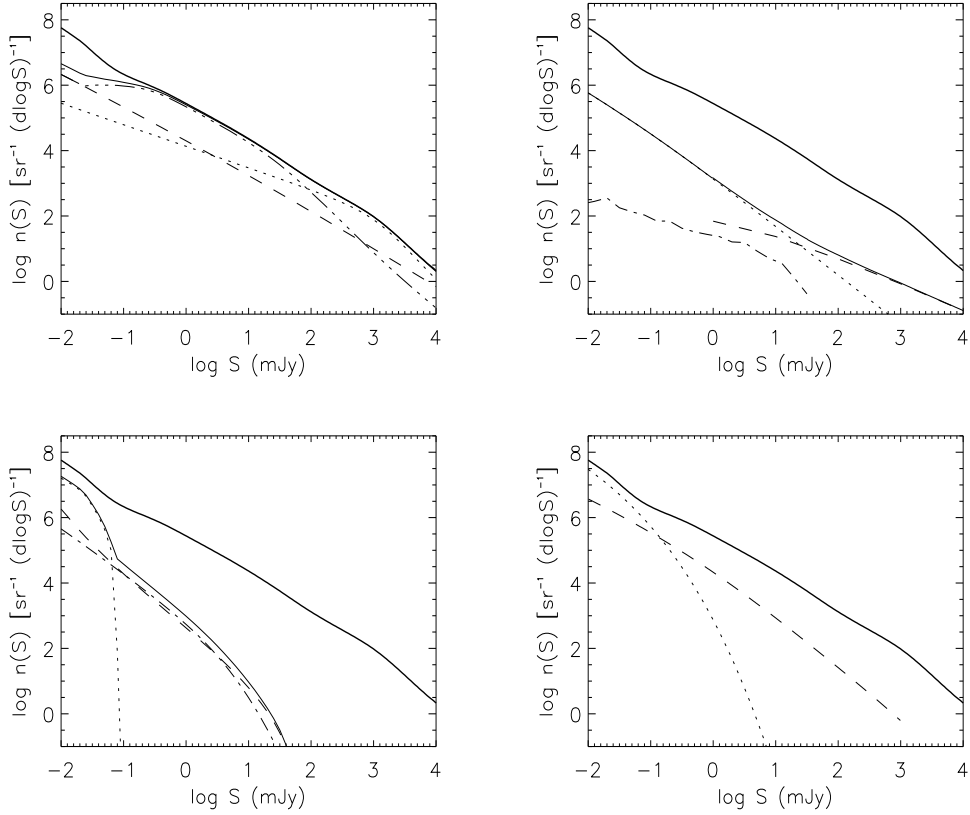}
\caption{Predicted differential counts at 20 GHz for various
extragalactic source populations. Upper left-hand panel (classical
radio sources): FSRQs (dotted line); BL Lacs (dashed line);
steep-spectrum sources (triple dot-dashed line). Upper right-hand
panel (special sources): ADAFs (dotted line); extreme GPS quasars
and galaxies (dashed line); GRB afterglows (dot-dashed line).
Lower left-hand panel (star-forming galaxies): proto-spheroids
(dotted line); spirals (dot-dashed line); starburst galaxies
(dashed line). Lower right-hand panel: SZ effects on galactic
scales (dotted line) and on cluster scales (dashed line). The sum
of contributions shown in each panel and the overall total counts
are indicated by a thin and thick solid line, respectively. }
\label{tot_counts20}
\end{center}
\end{figure*}

\begin{figure*}
\begin{center}
\includegraphics[height=11.5cm, width=14cm]{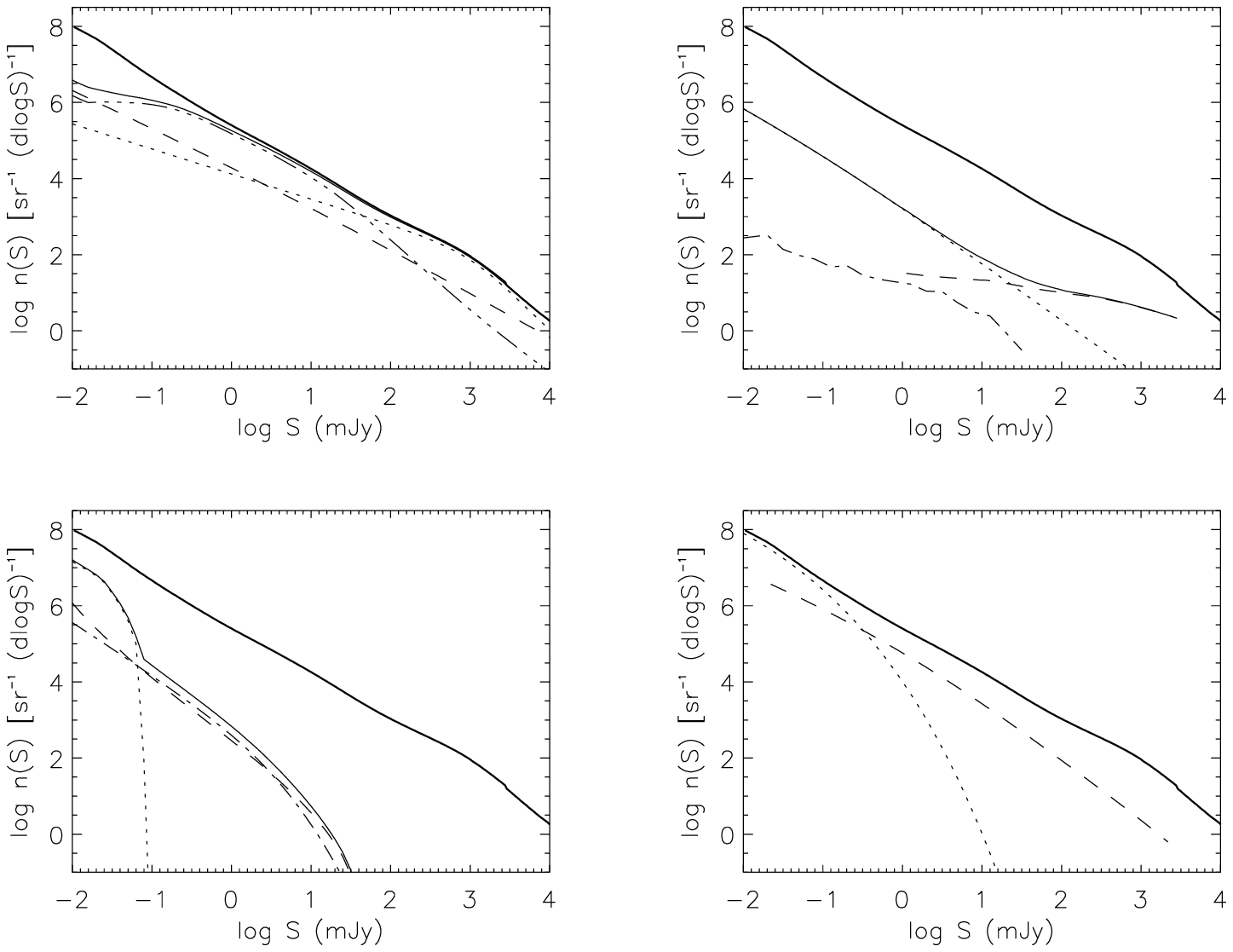}
\caption{Same as in Fig.~\protect\ref{tot_counts20}, but at 30
GHz. } \label{tot_counts30}
\end{center}
\end{figure*}

The afterglow emission of GRBs can be modelled as synchrotron
emission from a decelerating blast wave in an ambient medium,
plausibly the interstellar medium of the host galaxy (Waxman 1997;
Wijers \& Galama 1999; Meszaros 1999). The radio flux, above the
self-absorption break occurring at $\lsim 5\,$GHz, is proportional
to $\nu^{1/3}$ up to a peak frequency that decreases with time. In
the upper right-hand panel of Figs.~\ref{tot_counts20} and
\ref{tot_counts30} we show (dot-dashed line) the estimates of the
counts of GRB afterglows (kindly provided by B. Ciardi) yielded by
the Ciardi \& Loeb (2000) model. According to these estimates, a
large area survey at $\simeq 1\,$cm to a flux limit $\simeq
1\,$mJy should discover some GRBs (see also Seaton \& Partridge
2001).

%%%%%%%%%%%%%%%%%%%%%%%%%%%

\section{Discussion and conclusions}

A synoptic view of the contributions of the above source
populations to the 20 and 30 GHz differential counts is presented
in Figs.~\ref{tot_counts20} and \ref{tot_counts30}. As expected,
and directly demonstrated by the high frequency 9C (Taylor et al.
2001; Waldram et al. 2003) and ATCA (Ricci et al. 2004) surveys,
the dominant population at bright flux densities are blazars.
These are, however, a composite population, including the rapidly
evolving Flat Spectrum Radio Quasars and the slowly evolving BL
Lac objects. The former sub-population is prominent above $\simeq
100\,$mJy, but its counts converge rapidly at fainter flux
densities, eventually falling below those of steep-spectrum
sources and approaching those of BL Lacs.

The Figs.~\ref{tot_counts20} and \ref{tot_counts30} also show that
ongoing high frequency surveys are not far from serendipitous
observations of cluster SZ effects, easily recognizable because
they show up as negative peaks.

It is well known very deep surveys at 1.4 -- 8.4 GHz that a large
fraction of micro-Jy radio sources are moderate redshift active
star forming galaxies (Gruppioni et al. 2001; Fomalont et al.
2002; Chapman et al. 2003). Our models are consistent with these
results, and indicate that below a few tens of $\mu$Jy (at 20--30
GHz), there is a further change in the composition of the
radio-source population as proto-spheroidal galaxies come into
play, suddenly increasing the ratio of star forming to classical
radio sources. Since these objects are predicted to be at typical
redshifts of 2--3 and to be highly obscured by dust, there
appearance will also increase the fraction of optically very faint
sources and strongly shift to higher redshifts the redshift
distribution.

The emission of star forming galaxies in the spectral region of
interest here is dominated by synchrotron and free-free processes
at rest-frame wavelengths longer than a few mm, and by thermal
dust radiation at shorter wavelengths. The latter emission is
characterized by a steeply inverted spectrum, $S_\nu \propto
\nu^\alpha$, with $\alpha \sim 4$, causing a strongly negative
K-correction and a sharp steepening of the counts. Counts of
proto-spheroidal galaxies could be significantly higher than shown
in Figs.~\ref{tot_counts20} and \ref{tot_counts30} if the mm
excess detected in our own Galaxy, combining Archeops with WMAP
and DIRBE data (Bernard et al. 2003; Dupac et al. 2003), and in
NGC1569 is a general property of the SED of dusty galaxies.

Below $\simeq 100\,\mu$Jy the 20--$30\,$GHz a significant
contribution to the counts may come from Sunyaev-Zeldovich (1972)
signals (De Zotti et al. 2004; Rosa-Gonzalez et al. 2004) or
free-free emission (Oh 1999) produced by proto-galactic plasma.
While no attempts to estimate the counts of proto-galactic
free-free sources have been produced yet, we show in
Figs.~\ref{tot_counts20} and \ref{tot_counts30} the tentative
estimates of the counts of SZ effects by De Zotti et al. (2004),
that, as pointed out by the authors, should be more conservatively
interpreted as (possibly generous) upper limits.

\begin{acknowledgements}
We thank Benedetta Ciardi for having kindly provided the counts of
radio afterglows of $\gamma$-ray bursts. This research was partly
supported by the Italian Space Agency (ASI) and by the Italian
MIUR through a COFIN grant. DM acknowledges partial financial
support from the Fondazione ``G. Tovini'', and  LT and JGN from
the Spanish MEC, project ESP2002-04141-C03-01.

\end{acknowledgements}

\end{document}